# A Paradigm for the Application of Cloud Computing in Mobile Intelligent Tutoring Systems

Hossein Movafegh Ghadirli[1] and Maryam Rastgarpour[2]

[1] Graduate student in Computer Engineering, Young Researchers Club, Islamshahr Branch, Islamic Azad University, Islamshahr, Iran
`h.m.ghadirli@gmail.com; hossein.movafegh@iau-saveh.ac.ir`
[2] Faculty of Computer Engineering, Department of Computer, Saveh Branch, Islamic Azad University, Saveh, Iran
`m.rastgarpour@iau-saveh.ac.ir; m.rastgarpour@gmail.com`

## ABSTRACT

*Nowadays, with the rapid growth of cloud computing, many industries are going to move their computing activities to clouds. Researchers of virtual learning are also looking for the ways to use clouds through mobile platforms. This paper offers a model to accompany the benefits of "Mobile Intelligent Learning" technology and "Cloud Computing". The architecture of purposed system is based on multi-layer architecture of Mobile Cloud Computing. Despite the existing challenges, the system has increased the life of mobile device battery. It will raise working memory capacity and processing capacity of the educational system in addition to the greater advantage of the educational system. The proposed system allows the users to enjoy an intelligent learning every-time and every-where, reduces training costs and hardware dependency, and increases consistency, efficiency, and data reliability.*

## KEYWORDS

*Mobile Services, Cloud Computing, Mobile Intelligent Learning, Expert System*

## 1. INTRODUCTION

Nowadays growth of technology is fast and unpredictable in the economy, industry and personal issues [1]. One of the aspects of social life is the process of learning in universities, schools and other educational institutions. Extensive researches and huge investments have been carried out to develop technological learning in recent years. Now the word "Learning" is accompanied with the concepts such as *Electronic*, *Cognitive*, *Intelligent*, *Distance* and *Web based*. Since one of the attractive, efficient and widely used technologies is the use of mobile devices to do the tasks, researchers have tried to replace the previous notions with mobile learning. They develop educational softwares that can be implemented on mobile devices.

Mobile learning means the use of learning applications on mobile devices such as smart phones, PDA and tablets (unlike mobile devices which are small, portable, compact and pocket sized, Laptops are not considered as mobile systems, since they are expensive and heavy and they consume much energy) [2]. Recent researches indicate that the variety of learners, the training and learning process and infrastructure changes to subscribers, in addition to significant impact on learning quality, is more motivating learners. It causes wider interest of investors toward these softwares.

At the present, mobile devices are increasing rapidly, since they are the easiest and the most effective communication tools. In addition, their crucial role in human life, when and where to use them are not restricted (called ETEW[1]) [3], [4], [5]. Mobile users can use different applications on their devices or receive even different kinds of services through wireless networks distantly.

With increasing propagation of mobile devices technology, the popularity of this device has also increased. Some features such as mobility, optimized and easy to use are of the benefits of mobile devices. Nevertheless, the challenges of the resources of mobile devices (such as short battery life, small memory capacity and low bandwidth) and also of communication (such as mobility and data security) are the reasons for the decrease of service quality.

Cloud computing has been known as the Infrastructure of the next generation [4]. Cloud computing provides users with a way to share distributed resources and services of organizations in a cloud, and a platform and software is provided as a service in that infrastructure [7]. Cloud computing can present benefits for the users in the use of the infrastructures (such as servers, networks and storages), platform (such as firm-wares and operating systems), and softwares (such as applications) with a little cost. In addition, cloud computing providers (such as Google, Amazon, IBM, Sun Microsystems, Microsoft, IBM, and Sales-force) can use their resources flexibly, depending on the demands of the users [4].

Many educational institutions such as universities and schools would like to use software that can be hosted on the cloud; since it allows the final user (such as the softwares on his/her PC) needs no License, installation and maintenance of the softwares [8], [9]. In this regard, some cloud providers like Amazon, Google, Yahoo, Microsoft, etc. also support free hosting of e-learning systems [4]. Thus, this paper tries to present a paradigm for the application of cloud computing in mobile intelligent tutoring systems.

### 1.1. Related Works

In 2009 a system was introduced that provided private and virtual education for learners with regard to pedagogical rules [10]. But researchers were to transfer the complicate educational systems from PCs to mobile devices. The benefits of cloud computing and mobile learning integration have been pointed out in [11], one of which is increasing the quality of communication between the learner and the teacher. But they are mentioned in detail in section 4.2.

Some mobile applications already extract and aggregate information from multiple phones. Tweetie Atebits for the iPhone uses locations from other phones running the application to allow users to see recent Twitter posts by nearby users [12]. Video and photo publishing applications such as YouTube and Flickr allow users to upload multimedia data to share online. The Ocarina application Smule for the iPhone allows users to listen to songs played by other users of the application, displaying the location of each user on a globe. Such smartphone applications are "push"-based and centralized, meaning that users push their information to a remote server where it is processed and shared [12].

*Cornucopia* is one of the implemented examples of the proposed system, designed for the research affairs of undergraduate Genetic learners, and *Plantations Pathfinder* which was also designed to provide information for them, qua farms and gardens information were shown on mobile devices for visitors [13].

Another example of the system was presented in [14] that teaches some courses on image/video processing; using a mobile phone, learners are able to compare a variety of algorithms such as deblurring, denoising, face detection and image enhancement used in mobile applications.

The rest of this paper is organized as follows. Section 2 investigates mobile intelligent learning systems and its challenges. It also explains cloud computing and its derivative namely mobile

---

[1] Every Time and Every Where

cloud computing. Section 3 presents the proposed system in this paper and discusses its architecture. System evaluation is carried out in detail in section 5. Finally, the paper concludes in section 5.

## 2. MATERIALS AND METHODS

Cloud computing is not only related to personal computers, it also affects and heavily impact the mobile technology. In Mobile Cloud Computing both the data storage and the data processing happen outside of the mobile device i.e. when we combined concept of cloud computing in mobile environment. In Mobile Cloud Computing scenario all the computing power and data storage move into the mobile cloud[15]. In fact Cloud has generated many resources which can be used by various educational institutions and streams where their existing/proposed web based learning systems can be implemented at low cost.

### 2.1 Mobile Intelligent Learning System

Since 1980 that the use of computers began in learning process [16], the researchers have so far tried to make the educational systems more effective and easier. With the emergence of the phenomenon of AI now few systems can be found that does not use the minimum intelligence; In this context, the idea of integrating "intelligence" feature and static e-learning systems was also formed that resulted in the increased effectiveness of these systems in users' speed, quality and amount of learning.

Another aspect of e-learning systems is learning easiness. Users are often interested in being trained in anytime and anyplace they wish[17].

Not very long lifetime passes from the e-learning web-based systems; nevertheless one of the main factors that led to the use of "mobile learning" instead of "web-based learning" is the learners' lack of access to a computer (connected to internet). Reports show that in 2005, in many schools, there is one computer for every learner and the lowest rate of "computers to learners" is about 1 to 3 [18]. Though these factors do not represent the majority of the school's situations, but a 9% increase is observed in the use of handheld computers since 2003 [19]. These factors show a positive trend in the use of handheld computers at schools and it allows learners to balance their use of technology at home and school.

Applying mobile devices, universities, schools and other educational establishments can provide conditions for the use of intelligent learning systems without financial resources and construction of computer labs [20]. Learner can easily have educational system on their mobile devices and transport them between home and school. Mobility lets these systems also be used outside of computer labs and classrooms. As a result, opportunities can also occur for learners to learn at home and in other locations. so, universities, schools, and shopping centers' administrators can share a number of mobile devices for learners instead of computer labs and expensive PCs.

Generally, the benefits of cloud computing in e-learning can be divided into four groups [21]:

- reducing the costs of using resources
- flexibility in the use of infrastructure
- increased availability
- the client is the end user

Intelligent Tutoring System will make a specific model of learner's knowledge and characteristics and this model will get perfect during the interaction between the system and the learner. This model is compared with the domain model in the system to determine an appropriate strategy for tutoring learner [3]. It should be noted that users are not always in a fixed location or it is not possible for them to have free access to the internet all the time. Therefore, researchers have taken advantage of the potential of mobile devices to enable data transfers on intelligent learning application systems [2] and create "mobile intelligent tutoring systems" facilities.

The main characteristics of a mobile intelligent tutoring system are portability and intelligence; however, these systems also have disadvantages compared to desktop-based systems. Mobile intelligent tutoring systems face some challenges such as "implementation difficulty", which are discussed in Table 1.

Table 1. Challenges of a Mobile Intelligent Tutoring System [22].

| Field | Description of challenge |
|---|---|
| Interface | - Small monitor (2-5 inch)<br>- Difficult design as a single window<br>- Limited data entry with a small keyboard |
| Application | - Using for a short time, from few seconds to few minutes<br>- Having the role of the client with no content of itself |
| Architecture | - Low memory capacity<br>- Needing a cellular network or wireless internet |

## 2.2 Cloud Computing

The NIST[2] defines *Cloud Computing* as follows [23]:

It is a model for enabling ubiquitous, convenient, on-demand network access to a shared pool of configurable computing resources (e.g., networks, servers, storage, applications, and services) that can be rapidly provisioned and released with minimal management effort or service provider interaction. Cloud computing is a new computational method in which the infrastructure, platform and software is provided as a service. Using different computing resources, members of cloud computing can easily solve their problems by a cloud, and it gives users a lot of flexibility.

In fact, cloud computing can remove the limitations of the network and its hardware components and attract a lot of attention through its services to many components (Figure 1).

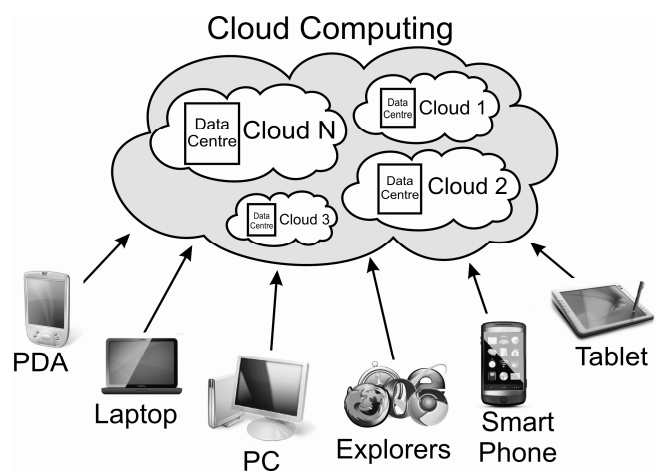

Figure 1. Cloud Computing Clients

---

[2] National Institute of Standards and Technology (NIST)

The great popularity of cloud computing is because of "computing" transfer; Instead of the local machine, the data center, located on the clouds, is responsible for computing task. So any device such as mobile phones, rather than doing difficult and complex calculations, will be able to send equation parameter to a service in a cloud and receive a quick response to it [24].

Cloud computing is generally a large-scale distributed network, which is implemented based on a number of servers (within the data center). This cloud model promotes availability and is composed of five essential characteristics, three service models, and four deployment models [23].

**2.2.1 Essential Characteristics**

1) *On-demand self-service.* A consumer can unilaterally provision computing capabilities, such as server time and network storage, as needed automatically without requiring human interaction with each service's provider.

2) *Broad network access.* Capabilities are available over the network and accessed through standard mechanisms that promote use by heterogeneous thin or thick client platforms (e.g., mobile phones, laptops, and PDAs).

3) *Resource pooling.* The provider's computing resources are pooled to serve multiple consumers using a multi-tenant model, with different physical and virtual resources dynamically assigned and reassigned according to consumer demand. There is a sense of location independence in that the customer generally has no control or knowledge over the exact location of the provided resources but may be able to specify location at a higher level of virtuality (e.g., country, state, or data center). Examples of resources include storage, processing, memory, network bandwidth, and virtual machines.

4) *Rapid elasticity.* Capabilities can be rapidly and elastically provisioned, in some cases automatically, to quickly scale out, and rapidly released to quickly scale in. To the consumer, the capabilities available for provisioning often appear to be unlimited and can be purchased in any quantity at any time.

5) *Measured Service.* Cloud systems automatically control and optimize resource use by leveraging a metering capability at some level of abstraction appropriate to the type of service (e.g., storage, processing, bandwidth, and active user accounts). Resource usage can be monitored, controlled, and reported, providing transparency for both the provider and consumer of the utilized service.[23]

**2.2.2 Service Models**

Cloud services, illustrated in Figure 2, are usually based on three layers [4]:

1) *Data Center* layer provides the required hardware and infrastructure of the clouds. In this layer, there are a number of servers connected to high-speed networks. Data centers are often located in places with the ability of high voltage power supply and away from any dangers.

2) *Infrastructure as a Service (IAAS)* is located on the data center which provides hardware, storage, servers and network components and the use these resources is based on users' needs; some examples of this layer are *Amazon Elastic Cloud Computing* and *Simple Storage Service(S3)*.

3) *Platform as a Service (PAAS)* is proposed as a developed environment for traditional software's building, testing and developing. Some examples of this layer are *Google App Engine*، *Microsoft Azure*, and *Amazon Map Reduce/Simple Storage Service*.

4) *Software as a Service (SAAS)* provides an application distribution with special needs. In

this layer, users can have access to their information and applications through internet and by paying for their own consumption. *Salesforce* is one of the pioneers in providing services in this way.

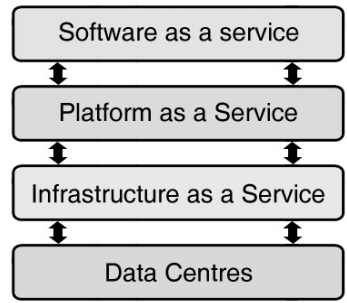

Figure 2. Architecture of Service-Oriented Cloud Computing [4]

## 2.3 Mobile Cloud Computing

In [25], mobile cloud computing (MCC) has been introduced as a new paradigm for mobile applications, in which, instead of running mobile software on mobile devices, it will be transferred to a centralized and powerful computing platform in the cloud. Simply "mobile cloud computing" refers to an infrastructure in which two operations "data storage" and "data processing" is done outside of the of mobile devices platform [26]. Centralized applications on clouds are available using a wireless connection.

Systems based on cloud computing, were introduced immediately after introducing mobile cloud computing technologies. Since these systems were able to reduce the development and implementation of mobile applications, so they attracted investor's attention as a profitable business; on the other hand, researchers mentioned it as a way to achieve *Green IT* [27].

MCC is a cloud service platform supporting many mobile application scenarios. Here, we just name a few: mobile health, mobile learning, mobile banking, intelligent transportation, smart grid/home, mobile advertising, urban sensing, disaster recovery, mobile entertaining/gaming, mobile social networks, and mobile enterprise solutions [28].

## 4. THE PROPOSED SYSTEM

The purpose of this study is to provide a model for integrating cloud technology with two components, *intelligence* and *mobile learning*, and a system derived from mobile cloud computing architecture.

In this model, instead of a powerful processor and large memory on their mobile devices, mobile users can use memory and processors in the clouds to run their programs. Any user connects to a private cloud environment, using his/her mobile device and username and password after authentication.

In a cloud environment, an intelligent learning system has been uploaded on one/several data centers and has saved a profile for each user. The mentioned learning system, using an expert system recognizes and offers the appropriate educational content based on user's talent, prior knowledge and characteristics [2].

The application of cloud computing to create virtual and private learning environments was welcomed by many institutions, since it reduced their costs and even sometimes made them free.

### 4.1. System Architecture

The proposed system architecture is shown in Figure 3. In this figure, mobile devices are connected to the Internet by the Base Transceiver Station. First, user's inquiry and data (such as user ID and location) are sent to the central processor. Then, "validation", "authorization" and "user accounts management" are carried out by mobile network operators.

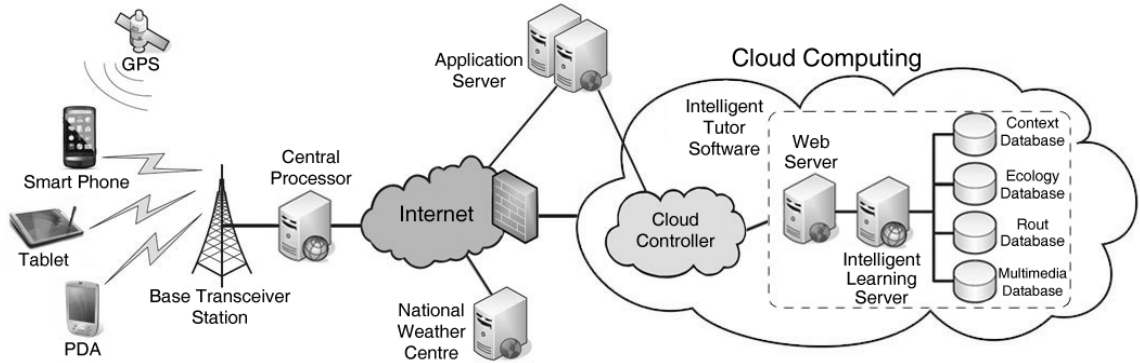

Figure 3. The Proposed System Architecture

User application is delivered to the cloud via internet. By processing the request, the cloud controller determines the appropriate service within the cloud. The service has been provided with the concepts "useful computing", "virtualization" and "Service Oriented Architecture".

Within the cloud, web server acts as an interface of Intelligent Learning Server and internet network. It also encodes the generated content so that it can be transferred by the web and displayed on mobile devices.

The intelligent learning server extracts educational content tailored to each user by an expert system and the existing information on the four existing databases [20].There are four main databases in this framework:

- *Concepts Database* keeps users' inherited characteristics.
- *Ecology Database* keeps features of mountains, beaches, vegetation, animals, insects, and so on.
- *Routing Database* keeps local and geographic tips with regard to the information of the *National Weather center* about data transfer routes.
- *Multimedia Database* save user-system interaction messages (e.g. text, audio, video and images).

The proposed system uses four-layer architecture as shown in Figure 2 and discussed in section 3. In order to demonstrate the impact of cloud models with user's needs, applying this architecture is very common [29].

## 5. SYSTEM EVALUATION AND DISCUSSION

This section explains the advantages and disadvantages of the proposed system for evaluation. In general, the advantages of the proposed system are given below:

- *The battery life of mobile devices increases*. Battery life is one of the main concerns of the users when they use mobile devices. When using intelligent learning systems on mobile devices, CPU, monitor, and memory spend a lot of time and electrical energy to do complex calculations, while using the cloud for processing and storage results in saving battery and helps the user to run the application faster. The conclusion of an evaluation [30] has shown that in distant processing, electrical power consumption is reduced to more than 45 percent.

- *Information storage space and processing capacity increases.* Limited space of memory on mobile devices doesn't allow users to store heavy educational content and profiles. But the data center of the cloud will provide an appropriate storage space for the mobile applications and enables them to manage the user and application information through a wireless connection. *Amazon Simple Storage Service* and *Image Exchange* are examples that provide a large storage space inside the clouds for mobile users [31]. Complicate applications require a lot of time and energy to run and, meanwhile, the hardware limitations of mobile devices prevent users to take advantage of these systems. So the proposed system, utilizing the clouds, reduces the cost of implementing these applications.

- *Reliability and Efficiency increases*. In order to increase reliability, data storage and running applications on clouds are more effective than mobile devices. Because, data and application are stored on several servers, and several back-up copies of the data are taken, as well.

- *Learning costs are reduced.* Cloud-based educational applications are considered free or low-cost way for teachers and learners since the data storage and processing is transmitted to a data center in the cloud [21].

- *File format compatibility is improved*. There is more consistency to open the files in this system.

- *It is not dependent on hardware*. In this system, if the user's mobile device changes, there will be no compromising in running programs and opening documents, and it will require no special hardware or software to buy, as well.

Thus, cloud computing is considered as a solution for mobile computing [4]. But despite the advantages of the proposed system, the system also faces challenges and disadvantages mentioned below:

Internet speed can affect the quality of learning.

For long-term training courses, it is more efficient to purchase a server and operate a data center than to utilize the clouds [24].

Learners' lack of skills in using mobile devices' educational systems has been considered as one of the disadvantages of the proposed system similar to the ref. [31].

The biggest concern of using clouds is their security, which has become a very important and critical issue [32], since both the software and its data are located on remote servers, and they may stop working or get crashed with no error display or even be attacked by hackers.

In September 2009, according to a behavioral study by *International Data Corporation (IDC)*, those basic challenges of cloud computing identified by organizations were determined, five of which are shown in Figure 4 [33].

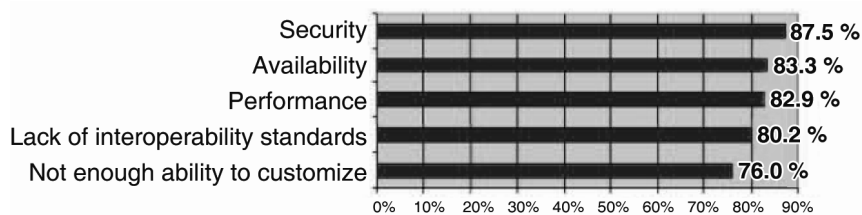

Figure 4.  Acceptable Challenges in Cloud Computing [32]

## 6. CONCLUSIONS

Mobile devices such as smart phones or tablets have a lot of popularity among users. This issue will pave the ground for the rise of mobile learning. The proposed model in this paper will be much appreciated in the future, because it is the result of the combined benefits of both mobile learning and cloud technologies. The applications can be run "distantly" and via mobile devices for the user in this model.

One of the major components of the proposed model, is the supercomputing which is responsible for computing and data storage required for mobile applications. This system provides a way to share resources and services among different users and helps to make learning for all users at any location possible. On the other hand, several companies are also able to share documents and files needed for training users within clouds.

Intelligent learning programs and data will be uploaded on the "data center" layer within the cloud. This system's architecture is based on multiple-layer architecture of mobile cloud computing. In systems implemented with the proposed model, the relationship between quality of service (QoS) and quality of experience (QoE) as a benchmark for measuring the performance of cloud-based systems is required. It has some valuable advantages as follows. It makes intelligent learning possible every-time and every-where. It can increase the battery life of mobile devices while using the educational system as well as raises the space of working memory and processing capacity of the education system. It also reduces learning costs and hardware dependency, and increases consistency, efficiency and reliability.

**Authors**

**Hossein Movafegh Ghadirli** received his B.S. in Computer Engineering from Saveh branch, Islamic Azad University (IAU), Saveh, Iran in 2009 and He is currently a graduate student in Computer Engineering at Science and Research branch, IAU, Saveh, Iran. His overriding interest has been bringing E-Learning, M-Learning and Intelligent Tutoring Systems to improve their productivity for both government and commercial organizations. He is a member of Young Researchers Club, Islamshahr Branch, Islamic Azad University, Islamshahr, Iran. 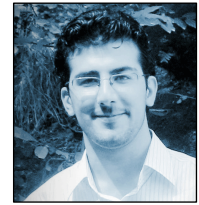

**Maryam Rastgarpour** received her B.S. in Computer Engineering from Kharazmi University, Tehran, Iran in 2003, and the M.S. in Computer Engineering from Science and Research branch, Islamic Azad University (IAU) , Tehran, Iran in 2007.She is currently a Ph.D. candidate in AI there. She is also a lecturer at Computer Department, Faculty of Engineering, Saveh branch, IAU for graduate and undergraduate students. Her research interests include in the areas of Machine Learning, Pattern Recognition, Expert Systems, E-Learning, Machine Vision, specifically in image segmentation and Intelligent Tutor System. 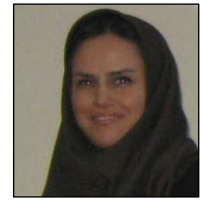